\documentclass[12pt,preprint]{aastex}
\shorttitle{Revised LHS Catalogue}
\shortauthors{Bakos, Sahu \& N\'emeth}

\begin{document}

\title{Revised Coordinates and Proper Motions of the Stars in the Luyten 
Half-Second Catalogue}

\author{G\'asp\'ar \'A.\ Bakos\altaffilmark{1,2,3}, 
Kailash C.\ Sahu\altaffilmark{1} and 
P\'eter N\'emeth\altaffilmark{4}\\
e-mail: gbakos@cfa.harvard.edu,ksahu@stsci.edu}
\altaffiltext{1}{Space Telescope Science Institute, 
3700 San Martin Drive, Baltimore, MD 21218, USA}
\altaffiltext{2}{Harvard-Smithsonian Center for Astrophysics, 
60 Garden Street, Cambridge, MA 02138, USA}
\altaffiltext{3}{Konkoly Observatory, P.O Box 67, H-1525 Budapest, Hungary}
\altaffiltext{4}{Department of Experimental Physics, JATE University,
Szeged, D\'om t\'er 9, H-6720 Szeged, Hungary}
 
\begin{abstract}

We present refined coordinates and proper motion data for the high
proper motion  (HPM) stars in the Luyten Half-Second (LHS) catalogue. 
The positional uncertainty in the original Luyten catalogue is
typically $>10\arcsec$ and is often $>30\arcsec$.  We have used the 
digital scans of the Palomar Observatory Sky Survey \mbox{(POSS) I} and
POSS II  plates to derive more accurate positions and proper motions of
the objects. Out of the 4470 candidates in the LHS catalogue, 4323
objects were manually re-identified in the \mbox{POSS I} and \mbox{POSS
II} scans. A small fraction of the stars were not found due to the lack
of finder charts and digitized \mbox{POSS II} scans. The uncertainties
in the revised positions are typically $\sim 2\arcsec$, but can be as
high as $\sim 8\arcsec$ in a few cases, which is a large improvement
over the original data. Cross-correlation with the Tycho-2 and
Hipparcos catalogues yielded 819 candidates (with $m_R \lesssim 12$). For these
brighter sources, the position and proper motion data were replaced
with the more accurate Tycho/Hipparcos data.  In total, we have revised
proper motion measurements and coordinates for 4040 stars and revised
coordinates for 4330 stars. In the printed version of the paper,
we present the updated coordinates and proper motion information  on
528 sources which represent the high proper motion subset  ($\mu >
1\arcsec$ yr$^{-1}$) of the LHS catalogue. The electronic version of
the paper{\footnote{The catalogue is available online at
http://www.stsci.edu/$\sim$ksahu/lhs}}  contains the updated
information on all the 4470 stars in the LHS  catalogue.

\end{abstract}

\keywords{astronomical data bases: high proper motion -- catalogues}

\section{Introduction}

High proper motion stars serve as useful probes for the determination
of many fundamental parameters, such as the stellar luminosity
function, luminosities and masses of individual stars, and the
structure and kinematics of the Galaxy. Among the few high-proper
motion catalogues available so far, the most exhaustive ones are those
of Luyten, which cover both the southern and the northern hemispheres,
and that of Giclas et al.~which covers only the northern hemisphere.
The Lowell Proper Motion Survey by Giclas et al.~(1971) has 8991 stars
in the northern hemisphere, with $\mu > 0.26\arcsec$ yr$^{-1}$, where
$\mu$ is the proper motion. Luyten's catalogues can be mainly divided
into 2 parts: the NLTT Catalogue, which has 58,845 stars with $\mu >
0.18\arcsec$ yr$^{-1}$ both in the northern and the southern hemispheres
(Luyten, 1961; Luyten, 1980); and the Luyten Half-Second Catalogue
(hereafter referred to as the LHS catalogue) \--- the main subject of
this paper -- which has the higher proper motion subset of 4470 stars
with $\mu > 0.5\arcsec$ yr$^{-1}$ (Luyten, 1979).

As explained in more detail later, the positional information of the
stars in the LHS catalogue have generally large uncertainties, which
can be as high as several arcminutes. However, the LHS catalogue is
used as the basis for many different studies, including the luminosity
functions of the halo population in the solar neighborhood 
(e.g.~Dawson, 1986) and the nearby white dwarfs (e.g.~Oswalt \& Smith
1995). Many astronomical projects,  particularly the ones that need
follow-up observations, would greatly benefit from  more  accurate
positions of the high-proper motion stars. So we undertook the task of
deriving accurate positions and proper motions for these high proper
motion stars using the Digitized Sky Survey I (DSS) I and DSS II
images, which are the digitized versions of the first and second epoch
Palomar Observatory Sky Survey (POSS) plates\footnote{Based on
photographic data obtained using The UK Schmidt Telescope.  The UK
Schmidt Telescope was operated by the Royal Observatory Edinburgh, with
funding from the UK Science and Engineering Research Council, until
1988 June, and thereafter by the Anglo-Australian Observatory. The
Digitized Sky Survey images  were produced from these photographic data
at the Space Telescope Science Institute under US Government grant NAG
W-2166.}. The results are presented in this paper. 

The paper is arranged as follows: in \S 2 we describe the details of
the LHS catalogue, in \S 3 we describe our procedure for determining
more accurate positions and proper motions of these stars, in \S 4 we
provide the details of the  use of Tycho-2/Hipparcos catalogues for the
brighter stars, in \S 5 we give the accuracy of our new catalogue and
some overall statistics, and in \S 6 we outline some suggestions for
future work.  The actual catalogue is given in tabular form, an online
version of which is available through the WWW at 
http://www.stsci.edu/$\sim$ksahu/lhs. The electronic version of the
paper and the online catalog contain information on all the 4470 stars
in the LHS catalogue, with updated positions and proper-motions for
4040 stars, and only updated positions for 4340 stars. The printed
version contains information on 528 sources which represent the
high proper motion subset  ($\mu > 1\arcsec$ yr$^{-1}$) of the LHS
catalogue. The finding charts for these sources can be 
obtained using the Digitized Sky Survey server at STScI 
(http://archive.stsci.edu/cgi-bin/dss\_form), by providing
the coordinates appropriate for the epoch of the DSS observations.

\section{Accuracy of LHS Coordinates and Magnitudes}

Table 1 gives the details of the number of entries in various
proper motion bins in the LHS catalogue. It is worth noting that out of 
the total of 4470 stars, 40 are common proper motion binaries.

Most of the HPM stars in the LHS and the NLTT catalogues were detected
and catalogued through a massive effort by W.~J.~Luyten, which involved
blinking the plates taken at two epochs, either by hand or by an
automated machine (Luyten, 1979). This was done for 804 fields, and the
remaining 160 low galactic latitude fields could not be processed
because of high density of stars. As a result, the catalogue contains
fewer stars in the low-galactic latitude fields than in the
high-galactic latitude fields. Furthermore, the ESO plates (covering
area south of $-33^\circ$ declination) were not available. So the
density of HPM stars is smaller south of $-33^\circ$ (limit of the
Palomar Survey) compared to the northern region.

The positions of a small fraction ($\sim$ 10\%) of these HPM stars were
measured from the meridian circle observations, for which `absolute'
positions are given in the LHS catalogue.  For the remaining majority
of the stars, only the `relative' positions are given. As a result,
although the LHS catalogue gives the positions of the stars to an
accuracy of 1s (or 15$\arcsec$) in RA and  0.1$\arcmin$ in dec, the
positional uncertainty is often larger. The uncertainty amounts to
as much as several arcminutes in some cases,  as discussed in the next
section.

Apart from the positions and the proper motion information, the LHS
catalogue also contains the estimated magnitudes. For a majority of the
stars, both red (R) and blue (pg) magnitudes are given, as determined
from the plates. In the LHS catalogue, the number of stars with no red
magnitude is 7, and the number of stars with no blue magnitude is 163. 
The magnitude distribution of the stars is shown in Fig.~1 which shows
that the catalogue has a limiting (red) magnitude of about 18. For
reference, a straight-line fit to the points in the brighter bins is
plotted, which indicates that the catalogue is affected by
incompleteness beyond $m_R \sim 14$. 

We would like to emphasize that the overall precision of the proper
motions (the magnitude of the motion $\mu$, as well as its position
angle $\theta$) in the LHS catalogue is generally good, only the
positions have high uncertainties (more details in the following
sections).

\section{The Method of Manual Search}

Manual inspection of several candidates using finder charts of the LHS
Atlas (Luyten \& Albers, 1979) revealed that the position errors can
readily exceed 1 arcmin (see upper panel of Fig.~2). Such a large
positional uncertainty makes it difficult to use an existing catalogue
such as the Guide Star catalogue (GSC) to derive more accurate
positions of the candidates through cross-correlation. Indeed, we first
attempted to derive accurate coordinates through an automated approach,
by correlating the positions of the LHS stars with the sources in the
GSC after performing the appropriate coordinate-transformations to the
epoch of the GSC observations. However, the number of matching pairs
was small even with a search radius of $30\arcsec$, yielding accurate
positions for only a small number of sources. If the search radius was
made bigger, the chance of finding another random star in the field was
high, and hence the cross-correlation technique was not reliable. The
automated search was made even more difficult by numerous plate-flaws,
dense stellar fields, possible minor planets, double stars, etc.

In order to reliably identify the HPM stars in an existing catalogue or
image, it is not only necessary to make sure that the candidate lies
within a specified search radius, but it is also important to confirm
that the object has a  high proper motion and has a similar brightness
as specified in the original catalogue. The first and second epoch POSS
plates are ideally suited for this purpose since (i) they cover the
whole sky, (ii) the limiting magnitude of the plates makes {\it all}
the LHS stars readily visible, and (iii) there are observations at two
epochs so that the motion of the HPM stars can be readily identified by
a comparison of the first and second epoch images. Furthermore, these
digital scans were originally made at STScI and hence are locally
available to the authors, which makes the task easier.  An example is
shown in Fig.~3 where the two panels show the DSS I and DSS II images
of LHS 36. The size of each image is 10 \arcmin x 10 \arcmin, and the
epochs of observations are 1953.28 and 1995.15 for of DSS I and DSS II,
respectively.  The HPM star is easily seen because of its motion
between the two epochs. Since the source coordinates as determined from
the DSS images are accurate to $\sim 1\arcsec$, the coordinates at two
epochs can be used to derive more accurate positions and proper-motion
data for these HPM stars.

To identify the candidates with the greatest certainty, we performed
{\em manual} identification of all the 4470 stars using the \mbox{DSS
I} and \mbox{DSS II} images and the finder charts of the LHS Atlas
(Luyten \& Albers, 1979), with the help of our self-written,
\textsc{IRAF}-based \textsc{gluyfin}, \textsc{gluypossi}
scripts\footnote{All the scripts are available from the first author on
request by e-mail}. The procedure for the manual identification is
briefly described below.

The DSS images for the two epochs were first retrieved through an
automatic script. The sizes of the images were selected to be large
enough so that the candidates would be in the field even with
$\sim1\arcmin$ initial errors in the coordinates and after undergoing
the proper motions for $\sim 40$ years, but as small as possible in
order to achieve good resolution which is required for precise
astrometry. The size of the images for Luyten stars 1 to 100 ($\mu
>2\arcsec$ yr$^{-1}$) was $15\arcmin\times15\arcmin$, the size of the
images for stars 101-1000 ($2\arcsec$ yr$^{-1} > \mu >1\arcsec$
yr$^{-1}$) was $5\arcmin\times 5\arcmin$, and the size of the images
for stars 1001-6433 ($1\arcsec$ yr$^{-1} > \mu > 0.48\arcsec$
yr$^{-1}$) was $4\arcmin\times 4\arcmin$ (cf.~Table~1). 

\mbox{DSS I} charts were retrieved for {\em all} fields, but 
\mbox{DSS II} charts were not available for 644 coordinates out of the
4470 (marked with ``P'' in Table~2). Four \mbox{DSS I}
and twenty-one \mbox{DSS II} images were of poor quality (edge of the
plates), which were not usable at all (marked with ``1'' and ``2'',
respectively).

\bf If both DSS scans were available, \rm and both scans had no major
defects (3801 cases), then the procedure for determining the positions
of the HPM star was as follows. The \textsc{gluyfin} script was used to
display and blink the two frames, and the HPM star was conspicuous by
its shift. Manual centering with a cursor and subsequent two
dimensional Gaussian-profile fitting were performed for both frames,
yielding precise pixel coordinates of the star for the two epochs.
These pixel coordinates were transformed to astrometric positions using
the \textsc{stsdas/gasp} package, which uses the plate-constants stored
in the header. (The resulting positions are in the GSC system, the
details of which are given later.) Using the two positions determined
for the two epochs, the proper motion of the star, its position angle
and its extrapolated position for epoch 2000.0 were computed. These
results are presented in Table~2.

Profile fitting {\em sometimes} failed or produced incorrect
coordinates if the HPM star was saturated (477 cases \-- flagged
as ``s''), or if the star is merged with another star in one of the scans
(246 cases \--- flagged as ``m'').
If the fitted position was obviously off the centroid, which was often
caused by the diffraction spikes of a saturated star, 
the parameters were fine-tuned, and in extreme cases manual
centering was performed (305 saturated and 64 merged stars
were re-fitted \--- flagged as ``c''). Double stars were
looked up from the LHS catalogue, so as to correctly
identify the components (``d'').

Our proper motion determination ($\mu, \theta$) is sometimes uncertain,
mostly because the positional shift of the HPM star between the two
epochs was not sufficient to determine the proper motion, or because
the star was blended on one of the images (522 cases). In
such a case, the star was flagged (``b''), and proper-motion data from
Luyten was used to compute J2000.0 coordinates, still using our
coordinates as initial values. Since the identification of the HPM star
is secure, use of the DSS coordinates clearly provides a more accurate
position of the star. The same procedure was adopted for all the HPM
stars with smaller shift than $5\arcsec$ between the two DSS scans 
(``B'').

If the identification from the DSS plate was uncertain, the
identification was reconfirmed using the finding charts in the LHS
Atlas (flagged as ``i''). However, in some cases finding charts were
not available in the LHS Atlas, though they would have been needed for
secure identification (140 cases, marked as ``W'').

Finally, if the star was not found, the original coordinates and proper
motion of Luyten are listed in Table~2 (flagged as ``N'').

\bf If only one DSS image was available \rm or had acceptable quality,
stars were identified using the finder charts of the LHS Atlas.
Astrometry was carried out on the single frame available using the
\textsc{gluypossi} script, and coordinates for J2000.0 were computed
from the proper motion given by Luyten. If the \mbox{DSS II} image was
not available, the star was flagged as ``P''. It is worth noting that
no Luyten stars with IDs greater than 6000 have finding charts since
they were compiled from published data. In many cases, we could
identify the star even without a finder, particularly when the initial
coordinates were relatively good, and the star was bright and isolated.

\section{Correlation with the Tycho-2 and Hipparcos Catalogues}
\label{sec:tycho}

The Tycho-2 catalogue is an astrometric reference catalogue containing
positions and proper motions as well as two-color photometric data for
the $\sim 2.5$ million brightest stars in the sky (H{\o}g et al.~2000).
Tycho-2 is based on observations of the ESA Hipparcos satellite, and
supersedes the earlier Tycho-1 catalogue (H{\o}g et al.~1997) both in
the number of sources and in astrometric precision. The limiting
magnitude ($V\sim11.5$) of Tycho-2 allows cross-identification of only
the bright LHS stars,  which considerably improves the precision
compared to the manual method, especially when the source is saturated
in the DSS image. We cross-correlated our {\em refined} coordinates 
with the coordinates in the Tycho-2 catalogue (note that in few cases,
e.g.~when ~no DSS images were available, these were identical to the
original LHS positions). Tests showed that the number of detections
saturated at a critical distance of $8\arcsec$ between coordinates
(which is used as one of the selection criteria).

Inspection of histograms of the magnitude differences both in the
``blue'' (Tycho B and Luyten photographic) and the ``red'' (Tycho V and
Luyten red) bands showed that red magnitudes have a better correlation,
which can be expressed as: $m_{Tyc,V}-m_{LHS,red}=0.1^m\pm^{1.1}_{0.5}$
(median$\pm$maximum width of the distribution of the magnitude
difference).  Using the combined criteria of position and magnitude
differences, 720 entries were refined and substituted by Tycho data
(flagged as ``T''). Double stars were handled manually, so as to avoid
confusion. The Luyten photographic and red magnitudes were substituted
by Tycho B and V magnitudes. 

The Tycho-2 catalogue Supplement No.~1 lists stars that were published
in the Tycho-1 or Hipparcos catalogues (Perryman et al.~1997), but not
listed in Tycho-2. Some of these stars were excluded from Tycho-2 as
they were too bright for proper treatment in the data reduction. We
searched the Supplement catalogue by selecting candidates with proper
motion and red magnitude measurements, i.e.~only by selecting stars that
were previously published in the Hipparcos catalogue, but not
necessarily in Tycho-1. The same detection criteria as in the case of
the Tycho-2 yielded 99 candidates.

\section{The Revised Catalogue} 

\subsection{Overall Statistics} 

Out of the 4470 HPM stars in the LHS catalogue, 3801 had both DSS I and
DSS II images with acceptable quality. Through a manual search as
explained above,  a total of 4323 stars were identified reliably, 12
stars had uncertain identifications, and 135 stars were not found. New
proper motion values were determined for 3894 stars. After {\em
cross-correlation} with the Tycho-2 and Supplement (Hipparcos)
catalogues, six additional stars were identified which were previously
not found, two uncertain identifications were clarified and the 
coordinates and proper motions were improved for 819 stars (720 from
Tycho-2, 99 from Supplement). The final number of entries with new
$(\mu, \theta)$ and coordinates are 4040 and 4330, respectively.

\subsection{Astrometric Accuracy}

In case of {\em manual identification}, the uncertainty in the final
astrometric position of the HPM star is due to several factors: (i) the
error in determining the center of the PSF at each epoch (ii) the positional
error in the reference catalogue (iii) the error due to the (imperfect
knowledge of the) geometric distortion of the plate, and (iv) the error in
the determination of the magnitude and direction of the proper motion and
the consequent error in the extrapolation of the position to the epoch 2000.

The point-spread functions (PSFs) in the majority of the DSS images had
FWHM of $\sim7\arcsec$ (\mbox{DSS I}) and $\sim4-5\arcsec$ (\mbox{DSS
II}),  while the plate scales are $1.68\arcsec/$pixel and
$1\arcsec/$pixel, respectively. The error in our profile-fitting,
except for the saturated and merged stars, was less than $0.2$ pixel
($\sim 0.2\arcsec$), the typical error being about $0.05$ pixel ($\sim
0.05\arcsec$). When the Gaussian fit failed, manual re-fitting could
have an error of $1$ pixel ($\sim1\arcsec$).
 
The absolute astrometry at each epoch also depends on the accuracy of
the plate-constants stored in the headers of the digital scans, and
their systematic errors caused by the reference catalogues used. The
northern hemisphere reference system is based on the AGK3 catalogue
(Dritter Katalog der Astronomischen Gesellschaft), and southern
hemisphere reference system is based on SAOC (Smithsonian Astrophysical
Observatory Catalogue) in the region north of $-65^\circ$ and CPC (Cape
Photometric Catalogue for 1950.0) in the far south, below $-65^\circ$.
The positional accuracy of the northern reference catalogue is, in
general, 3 times better than the southern one ($0.6\arcsec$
vs.~$1.7\arcsec$).

Positional errors caused by the geometric distortion from the plate
center to the edge are in the range $0.5\arcsec$ to $1.1\arcsec$ in the
northern celestial hemisphere, and $1.0\arcsec$ to $1.6\arcsec$ in
the southern celestial hemisphere (Taff et al.~1990).

In order to determine the position of the HPM star for the epoch 2000.0, 
we need to extrapolate the position from the POSS epoch, using the derived 
magnitude and direction of its proper motion. This procedure accordingly
increases the uncertainty in the final astrometric position. 

We estimate that the combined effect of these uncertainties in the
final astrometric position would be typically 2 arcsec, but can be 
$\sim 8\arcsec$ in a few cases. This is confirmed by Fig.~4, which
shows that the difference between our coordinates and the Tycho
coordinates peaks at 2$\arcsec$, beyond which it drops rapidly and 
approaches zero beyond  $5\arcsec$. We also tried to empirically
estimate the final error by comparing the observed positions of a few
HPM stars (as given in the GSC) with the derived positions using our
method. We transformed the positions of several stars to the epoch of
the appropriate Guide Star Catalogue (GSC) fields (which are based on
Palomar Quick-V and the SERC-J survey), and compared them with the
position of the GSC star. The positions were consistent within
$\sim5\arcsec$, as expected. We note that the uncertainty  is dominated
by the systematics explained above, and not by the accuracy in the
determinations of the centroids of the stars at each epoch. So the
uncertainties are not expected to be correlated with the magnitudes of
the stars, which was further confirmed  by making plots similar to
Fig.~4 for stars in different magnitude bins.

The accuracy of the proper motions ($\mu$ and $\theta$) is more
difficult  to estimate since they clearly depend on the timespan
between the two epochs of observations, the quality of the images, etc.
But comparison of  the proper motions with the Tycho-2 catalog gives a
fair estimate, which is shown in Fig.~4. We estimate that for reliably
identified sources, the accuracy in $\mu$ is generally $\pm 0.1 \arcsec
yr^{-1}$, and the accuracy in $\theta$ is $\pm 5^\circ$. But such a
comparison with the Tycho-2 catalog is valid only for  the brighter
stars. However, we note that the DSS goes much deeper than the 
magnitude limit of the LHS catalog, and the predominant uncertainty in
the proper motions comes from the timespan  between the two epochs
rather than the brightness of the source.  So the error is not likely
to be larger than twice these values even for the fainter sources.

The accuracy is naturally much higher for brighter sources with  {\em
Tycho-2/Hipparcos} data. In these cases, the standard errors  in the
coordinates and the proper motions for all the stars are 60 mas and 2.5
mas yr$^{-1}$, respectively, but if $m_{Tyc,V}<9.0^m$, the errors in
the coordinates are less than 7 mas. Thus there is a large difference
between the precisions of the Tycho-2 entries and the entries made with
manual  identification.  In this sense, our catalogue in not
homogeneous; but we have provided the best measurements that are
currently available in all cases.

\section{Summary and Suggestions for Future Work} 

We have revised the coordinates and proper motion data for the high
proper motion (HPM) stars in the Luyten Half-Second (LHS) catalogue. 
The positional uncertainty in the original Luyten catalogue is
typically $>10\arcsec$ and is often $>30\arcsec$. The uncertainties in
the revised positions are typically $\sim 2\arcsec$, but can be as high
as $\sim 8\arcsec$ in a few cases. The accuracy in $\mu$ is generally
$\pm 0.1 \arcsec yr^{-1}$, and the accuracy in $\theta$ is $\pm
5^\circ$. Out of the 4470 candidates in the LHS catalogue,  we have
revised proper motion measurements and coordinates for 4040 stars and
revised coordinates for 4330 stars. For most of the brighter sources 
($m_R \lesssim 12$), the position and proper motion data have been
replaced with the more accurate Tycho/Hipparcos data.  

As described in \S 1, the LHS catalogue contains only a subset of the
HPM stars currently available in the literature. It would be useful if
the work presented here is extended to include the full set of the HPM
stars, including those in the Luyten catalogue of HPM stars with $\mu >
0.2\arcsec yr^{-1}$, for which the coordinates have large
uncertainties.  Indeed,  now that the DSS I and the DSS II images are
available for the whole sky, it should be possible to produce a
complete catalogue of all the HPM stars in the entire sky including the
Galactic plane region, down to stars with  $\mu < 0.1\arcsec yr^{-1}$.
Fortunately, such a project has been undertaken by a group at STScI,
and the product will be extremely useful for several projects. Such
projects would include, to name a few, (i)~the prediction of future
microlensing events of background stars by HPM stars, the observations
of which can be used to derive accurate masses of the HPM stars (see,
e.g.,~Salim and Gould 2001; Dominik and Sahu 2000; Paczy\'nski 1998); 
(ii)~cross-correlations with other catalogues to obtain data at other
wavelengths; and (iii)~determining the contributions of possible halo
populations in the solar neighborhood (e.g.~Schmidt 1975; Dawson
1986; Oppenheimer et al. 2001; Reid, Sahu and Hawley 2001). 

\acknowledgements{Most of this work was done while G\'AB was enjoying
the hospitality of Space Telescope Science Institute (operated by NASA
for AURA) as a summer student. P.~N\'emeth and G\'AB would like to
thank G.~F\H{u}r\'esz for providing computer facilities to the project.
This project was supported by the DDRF grant of STScI. We thank the
anonymous referee for useful suggestions.}

\clearpage

\clearpage
\begin{figure}
\plotone{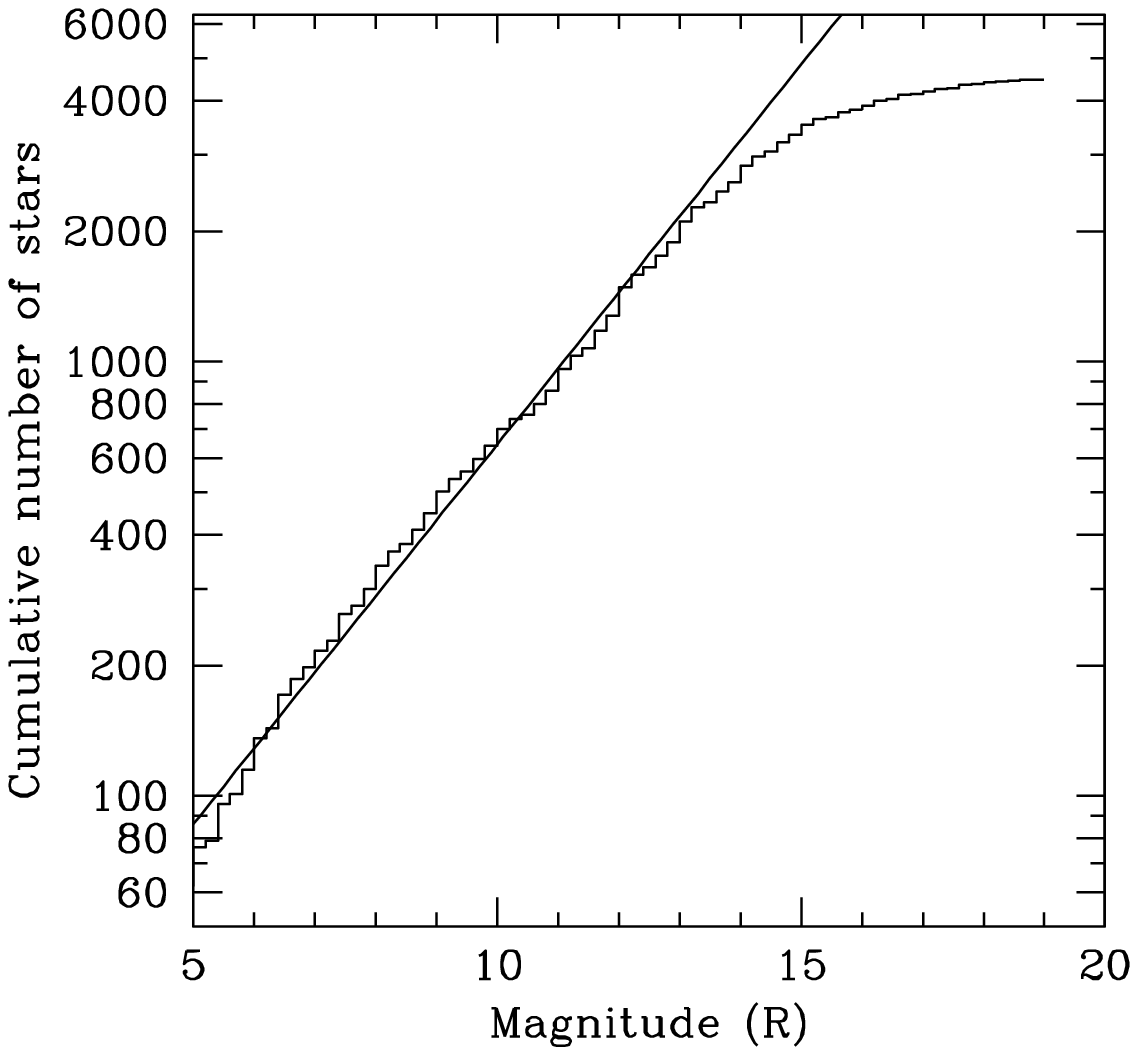}
\caption{The figure shows the cumulative number distribution of stars
as a function of magnitude. This shows that the LHS catalogue has a
limiting (red) magnitude of about 18. A straight line is shown for
reference which indicates that the catalogue is affected by
incompleteness beyond $m_R \sim 14$. }
\end{figure}
\clearpage
\begin{figure}
\plotone{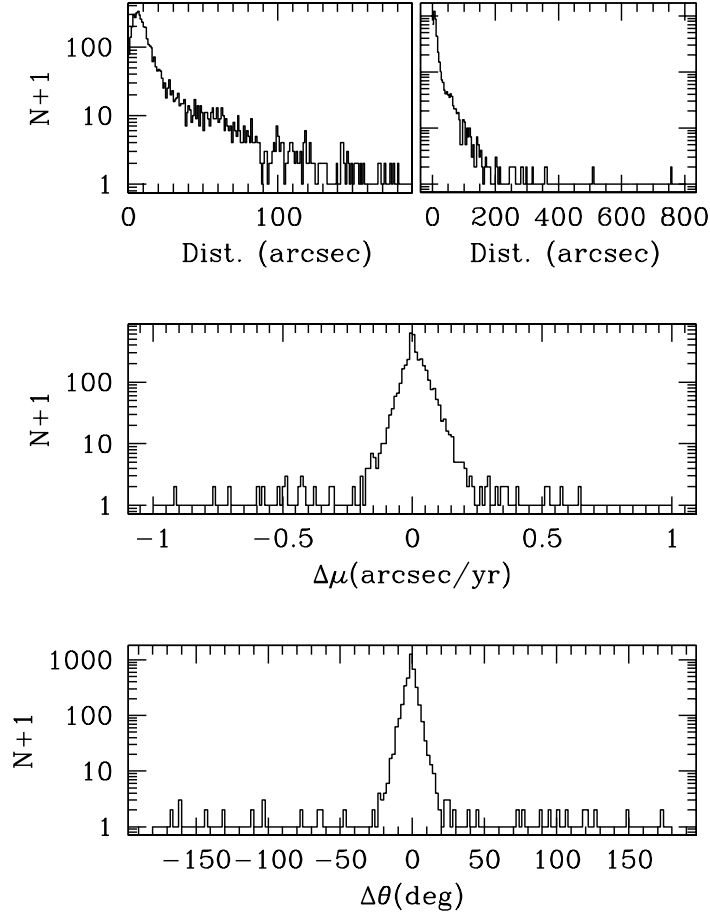}
\caption{The upper panel shows a histogram of the distance
between our positions (either result of the manual method or
cross-correlation with the Tycho catalogue) and the positions in the
LHS catalogue, both for epoch and equinox 2000.0. The upper left panel
uses $1\arcsec$ bins, while the upper right panel employs $4\arcsec$
bins. Only those stars were included, which were identified by the
manual search or cross-correlation with the Tycho catalogue. Note the long
tail of the distribution. The mid-panel shows the difference between
Luyten's and our proper motion in 200 equally spaced bins with
$0.01\arcsec$ binwidth. The lower panel displays the angle between the
star's motion determined by Luyten and by the present work (180 bins of
$2^\circ$ width). The lower two panels are shown only for stars not
flagged as ``B'' or ``b'' in Table~2, i.e.~when we have new
$\mu$ and $\theta$ measurements.}
\end{figure}
\clearpage
\begin{figure}
(These figures can be obtained from http://www.stsci.edu/$\sim$ksahu/lhs)
\caption{An example of the DSS I (left) and DSS II (right) 
images used for determining the coordinates and the proper motions. The
images shown here correspond to LHS 36, and the size of each image is
$10 \arcmin \times 10 \arcmin$. The epochs of observations are 1953.28 and
1995.15 for DSS I and DSS II, respectively. The HPM star is easily seen
because of its motion during the two epochs. North is up, and east is
to the left.}
\end{figure}
\clearpage
\begin{figure}
\epsscale{1.0}
\vspace*{-3.5cm}
\plotone{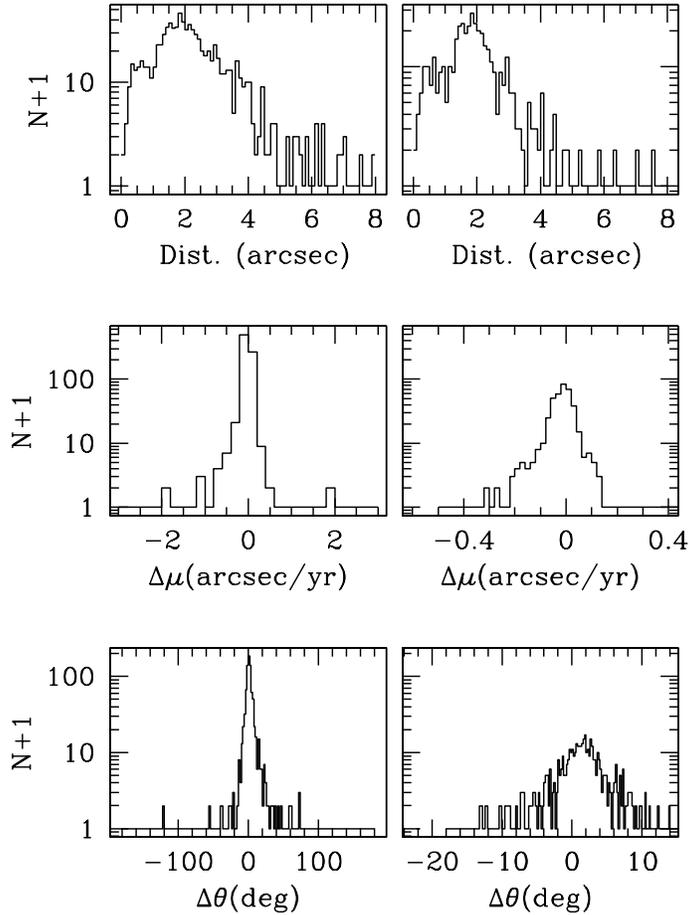}
\vspace*{-2.5cm}
\caption{The upper left panel shows a histogram of the distance
between the position as determined from the manual identification and
that of the Tycho catalogue for all the stars which were found both in
the Tycho catalogue and in the manual search (805 entries). The width
of the bins is $0.1\arcsec$. The upper right panel is almost the same,
but here the stars with ``B,b,s,i,m,c'' flags in Table~2,
i.e.~those with inaccurate proper motion measurements, saturated
profiles, etc., are not included. The two histograms are very similar,
which shows that  the uncertainty in the proper motion (and the
consequent uncertainty in the extrapolation to epoch 2000.0) is only a
second order effect compared to the original positional errors. The
middle panel shows the difference between the proper motions as
determined from the manual identification and that of the Tycho
catalogue. The middle left panel is for all stars (found both in the
Tycho catalogue and in the manual search), using $0.2\arcsec/yr$
binwidth. This shows that saturation of the star or inadequate timespan
between the two observations can yield very inaccurate proper motion
measurements. The middle right panel shows the same ($0.02\arcsec/yr$
binwidth, 384 stars), but after filtering all saturated and merged
stars, and those with inaccurate proper motion estimates
(``B,b,s,i,m,c'' flags). In these cases, our manual method has high
accuracy. The lower left panel displays the angle between the star's
motion in the Tycho catalogue and that of the manual method (180 bins
of $2^\circ$ width). The lower right panel is the same but after
filtering as in the previous panels. Again, this shows that the
accuracy in $\theta$ is high if the star is not  affected by
saturation or merging. }
\end{figure}

\begin{table*}
\begin{flushleft}
{\caption[]{Stars in the LHS catalogue}\label{tab:sum}}
\begin{tabular}{llcccccc}
\\
\hline
\\
Location in catalogue & Proper motion ($\mu$) & No.~of stars & \% of total & \\
&($\arcsec/yr$)\\
\hline
\\
Main body, 1-100 & $>$2 & 73 & 1.6 &\\
Main body, 101-1000 & $1 - 2$  & 455 & 10.2 &\\
Main body, 1001-5000 & $0.5 - 1$  & 3073 & 69 &\\
Appendix I, 5001-6000 & $0.48 - 0.499$  & 441 & 9.4 &\\
Appendix II, $>$6001 & $>0.49^*$  & 428 & 9.8 &\\
\\
All & & 4470 &  &\\  
\\
\hline
\end{tabular}
\end{flushleft}
{$^*$stars for which at one time or another a value of $\mu > 0.49$
$\arcsec$ yr$^{-1}$ was published}
\end{table*}

\clearpage
\pagestyle{empty}
\begin{deluxetable}{lrrrrrrrrrlrrrrrrrrrrrrr}
\rotate 
\tablewidth{0cm}
\setlength{\tabcolsep}{0.07in}
\tabletypesize{\tiny}
\tablecaption{Revised Positions and Proper Motions for LHS
stars\label{tab:data}}
\tablehead{
\multicolumn{1}{c}{LHS} &
\multicolumn{10}{c}{New data} &
\multicolumn{8}{c}{Luyten's original data} &
\multicolumn{2}{c}{Epoch}\\
\cline{3-9}
\cline{12-19}\\
\multicolumn{1}{c}{No.} &
\multicolumn{3}{c}{RA (J2000.0)\tablenotemark{a}} &
\multicolumn{3}{c}{Dec (J2000.0)\tablenotemark{a}} &
\multicolumn{1}{c}{$\mu$\tablenotemark{b}} &
\multicolumn{1}{c}{$\theta$\tablenotemark{b}} &
\multicolumn{1}{c}{Dist.\tablenotemark{c}} &
\multicolumn{1}{c}{Com.\tablenotemark{d}} &
\multicolumn{3}{c}{RA (J2000.0)} &
\multicolumn{3}{c}{Dec (J2000.0)} &
\colhead{$\mu$} &
\colhead{$\theta$} &
\colhead{POSS I} &
\colhead{POSS II} &
\colhead{$m_{red}$\tablenotemark{e}} &
\colhead{$m_{blue}$\tablenotemark{e}} &
\colhead{Tycho-2/HIP\tablenotemark{f}}}
\startdata
\input{data_ev.tex}
\enddata
\tablenotetext{a}{Equatorial coordinates for epoch and equinox 2000.0 using
our astrometry and proper motion measurements. If none of the digital scans
of POSS plates were available, or the difference between the epochs was not
sufficient, proper motions from Luyten were accepted. If star was not found,
both coordinates and proper motions were taken from Luyten.}
\tablenotetext{b}{$\theta$ is in degrees, $\mu$ is in $\arcsec/yr.$}
\tablenotetext{c}{Distance between predicted J2000.0 position of LHS and our
measurements in arcseconds. If coordinates of the LHS Catalogue were 
accepted, the distance is flagged as a ``\--\--'' dash.}
\tablenotetext{d}{Comments are the following:
1 \--- DSS I image has poor quality,
2 \--- DSS II image has poor quality,
B \--- position shift between the two frames is less than $5\arcsec$, proper
motion data from LHS Catalogue is taken,
H \--- coordinates, proper motion and magnitudes are from the Tycho-2
Supplement-1 catalogue, i.e.~from the Hipparcos catalogue,
N \--- star was not found in {\em manual} search,
P \--- DSS II frame was not available,
W \--- finder chart would have been needed for {\em manual} identification,
but was not available, 
T \--- coordinates, proper motion and magnitudes are from the Tycho-2
catalogue,
b \--- for some reason our proper motion measurement is not accurate,
Luyten's data is used,
c \--- refitting of position was done (due to saturation or merging),
d \--- double star, or companion of a binary,
m \--- merging,
i \--- identification dubious,
s \--- saturated on at least one of the frames
}
\tablenotetext{e}{Red and blue magnitudes are Tycho (Hipparcos) V and B if
star was found in the Tycho-2 (Supplement-1) catalogue, otherwise LHS red
and photographic magnitudes.}
\tablenotetext{f}{Identification of Tycho-2 catalogue is of the format
``tyc1-tyc2-tyc3'', while Hipparcos is the single Hipparcos number followed
by the CCDM component identifier (Dommanget \& Nys 1994).}
\tablecomments{Table with refined positions for Luyten stars. 
The printed version of the paper contains only the high proper motion 
subset ($\mu> 1\arcsec$ yr$^{-1}$) of the revised LHS Catalogue.
The electronic version of the paper contains the unabridged version
of the catalogue, which can also
be retrieved from and http://www.archive.stsci.edu/$\sim$ksahu/lhs.}
\end{deluxetable}

\end{document}